\documentclass[twocolumn,aps,amsmath,amssymb,floatfix,showpacs,superscriptaddress]{revtex4}
\usepackage{amsmath,amssymb,eucal,graphicx,bm}
\begin{document}
\title{Discrete Analog of the Burgers Equation}
\author{E.~Ben-Naim}
\affiliation{Theoretical
Division and Center for Nonlinear Studies, Los Alamos National
Laboratory, Los Alamos, NM 87545, USA}
\author{P.~L.~Krapivsky}
\affiliation{Department of Physics, Boston University, Boston, MA 02215, USA}
\begin{abstract}
We propose the set of coupled ordinary differential equations
$dn_j/dt=n_{j-1}^2-n_j^2$ as a discrete analog of the classic Burgers
equation.  We focus on traveling waves and triangular waves, and find
that these special solutions of the discrete system capture major
features of their continuous counterpart.  In particular, the
propagation velocity of a traveling wave and the shape of a triangular
wave match the continuous behavior. However, there are some subtle
differences. For traveling waves, the propagating front can be
extremely sharp as it exhibits double exponential decay. For
triangular waves, there is an unexpected logarithmic shift in the
location of the front. We establish these results using asymptotic
analysis, heuristic arguments, and direct numerical integration.
\end{abstract}       
\maketitle

The classic Burgers equation
\begin{equation}
\label{be}
n_t + (n^2)_x = \nu n_{xx}
\end{equation}
is the simplest partial differential equation which incorporates both
nonlinear advection and diffusive spreading
\cite{Waves,Logan,LL87,B96,book}.  This ubiquitous equation emerges
naturally in the presence of dissipation, and it is broadly used to
model traffic flows \cite{Waves}, transport processes \cite{szia,bd},
surface growth \cite{kpz,bs}, and large scale formation of matter in
the Universe \cite{yz,sz}.

The Burgers equation has two important properties.  The first is
continuity: equation \eqref{be} can be written in the form $n_t +
J_x=0$, hence assuring mass conservation in the absence of sources or
sinks.  If we view the quantity $n$ as a density, then the total mass
is a conserved quantity, $\int_{-\infty}^\infty dx\,n(x,t)={\rm
  const.}$, as long as the density vanishes, $n(x)\to 0$ in the
limits $x\to\pm \infty$. The second property is asymmetry: due to the
nonlinear advection term, equation \eqref{be} is not invariant with
respect to the inversion transformation $x\to -x$.

Our goal is to construct a discrete (in space) counterpart of the
Burgers equation that maintains these two properties. Thus we
discretize the spatial coordinate but keep the time variable
continuous, $n(x,t)\to n_j(t)$, where $j$ is integer.  The 
differential equation
\begin{equation}
\label{continuity}
\frac{dn_j}{dt} = f(n_{j-1}) - f(n_j)
\end{equation}
represents a continuity equation on a one-dimensional lattice. Indeed,
a finite total mass $M=\sum_{j=-\infty}^\infty n_j$ remains constant,
$M=\text{const.}$, if two conditions are met: (i) a vanishing density
$n_j\to 0$ as $j\to \pm\infty$, and (ii) a vanishing function $f$ at
the origin, $f(0)=0$.  To reproduce the nonlinear advection term in
\eqref{be}, we take a purely quadratic and positive function
$f(n)=n^2$. With this choice, we arrive at the set of nonlinear
difference-differential equations
\begin{equation}
\label{dbe}
\frac{dn_j}{dt}=n_{j-1}^2-n_j^2.
\end{equation}
This system of equations meets the two criteria of mass
conservation and asymmetry.

Immediately, we can point out an important difference between the
discrete equation \eqref{dbe} and the continuous equation \eqref{be}.
Let us treat the spatial variable in \eqref{dbe} as continuous, $j\to
x$, and replace the difference with a second order Taylor expansion.
The result of these two steps is the continuous equation
\begin{equation}
\label{cdbe}
n_t + (n^2)_x = (nn_x)_{x}.
\end{equation}
By construction, the nonlinear advection term is the same as in
\eqref{be}. However, the viscosity equals the density, $\nu=n$,
whereas in the original Burgers equation, the viscosity is constant.
We restrict our attention to positive densities, 
\begin{equation}
\label{positive}
n > 0,
\end{equation}
so that the solutions of \eqref{dbe} are stable (avoiding a negative
diffusion instability). We note that transport coefficients often
depend on density or temperature; in fluid dynamics \cite{LL87,B96},
for example, transport coefficients vary as $\sqrt{T}$ for hard-sphere
gases.

Equation \eqref{dbe} describes the evolution of the probability
density in a two-body analog of the standard Poisson process
\cite{poisson}. For example, we mention a homophilic network growth
process \cite{mejn}. In the canonical random network model, a pair of
nodes are chosen at random and subsequently, the two are connected by
a link. This elementary step is repeated indefinitely, and in finite
time, a percolating network emerges.  As a model of homophilic
networks where only similar entities interact, we considered the
situation where only nodes with exactly the same degree can be
connected \cite{bk}.  The degree distribution $n_j(t)$, that is, the
fraction of nodes of degree $j$ at time $t$, obeys the rate equation
\eqref{dbe}.  The initial condition $n_j(0)=\delta_{j,0}$ represents a
disconnected set of nodes.  In this network context, the quantity
$n_j(t)$ is a probability density, and mass conservation guarantees
proper normalization, $\sum_{j=0}^\infty n_j=1$. Moreover, the
condition \eqref{positive} reflects that probability distribution
functions are by definition positive.

In this paper, we discuss the solutions of the discrete equation
\eqref{dbe} in view of the well known solutions of the continuous
equation \eqref{be}. Using a combination of theoretical and numerical
methods, we analyzed the solutions of the discrete equation for the
following standard initial conditions \cite{Waves}: (i) a step
function resulting in a traveling wave, (ii) a localized delta
function leading to a triangular wave, and (iii) a complementary step
function with an ensuing rarefaction wave.  For all of these cases, we
find that the discrete analog faithfully captures the primary features
of the continuous Burgers equation. However, we also find subtle and
interesting departures from the classical solutions in the first two
cases.  Therefore, in the rest of this paper, we focus on traveling
waves and triangular waves.

\section*{Traveling Waves}

\begin{figure}[t]
\includegraphics[width=0.45\textwidth]{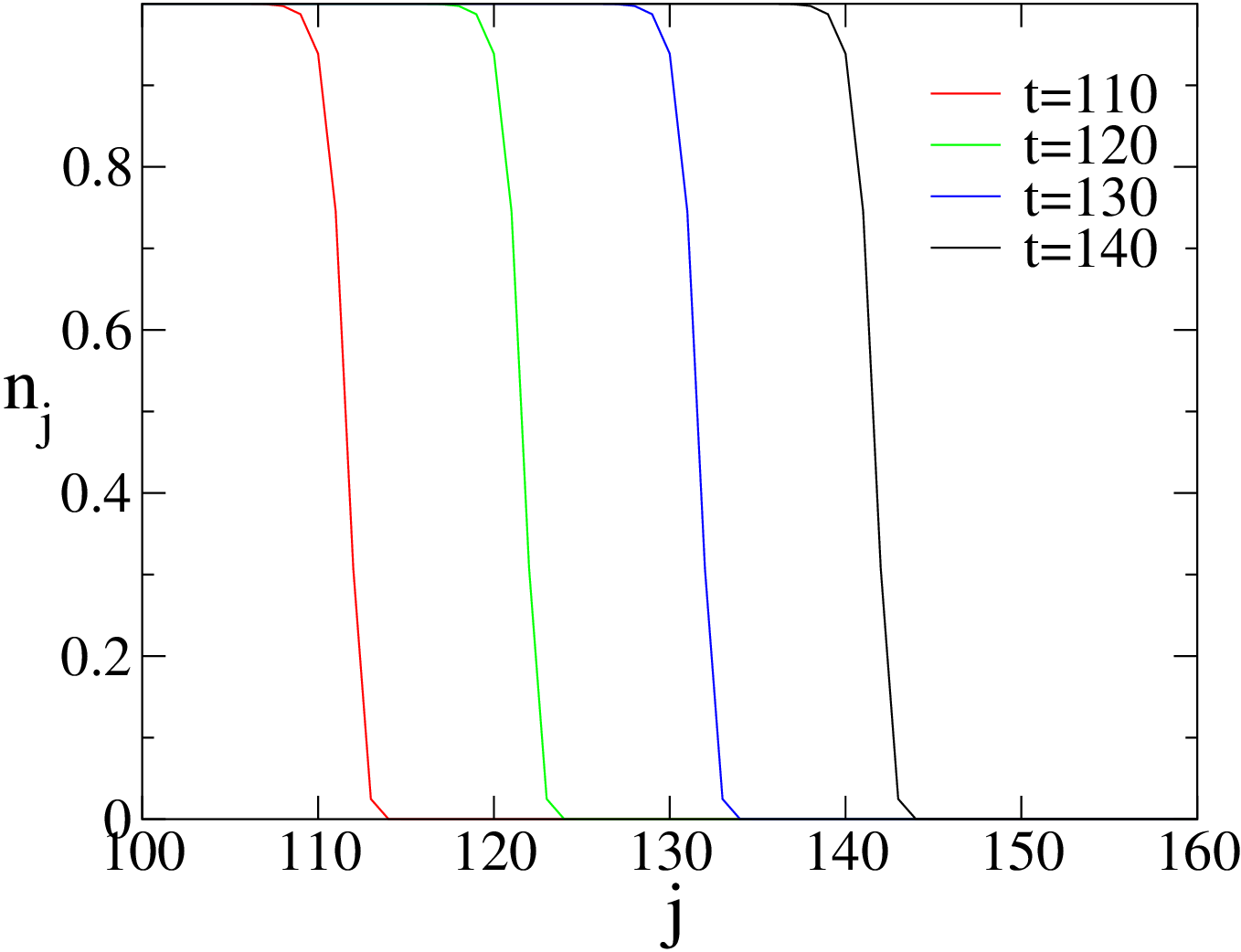}
\caption{The traveling wave. Shown is the function $n_j(t)$ versus $j$
at four different times: $t=110$, $120$, $130$, $140$. The results are
from numerical integration of the Eq.~\eqref{dbe} subject to the step
initial condition \eqref{step}.}
\label{fig-traveling-wave}
\end{figure}

We first consider the step function
initial condition
\begin{equation}
\label{step}
n_j(0)=
\begin{cases}
1 & \quad j\leq 0\\
0 & \quad j>0.
\end{cases}
\end{equation}
Using the convenient Adams-Bashforth method \cite{ba,dz}, we
numerically integrate the rate equation \eqref{dbe} and find that the
solution approaches a traveling wave (figure
\ref{fig-traveling-wave})
\begin{equation}
\label{wave}
n_j(t)\to G(j-vt),
\end{equation}
in the long-time limit. The function $G(z)$ specifies the form of the
traveling wave and $v$ is the propagation velocity. Of course, the
function $G(z)$ has the limiting behaviors $\lim_{z\to -\infty}G(z)=1$
and $\lim_{z\to \infty}G(z)=0$.

To find the propagation velocity, we first note that according to
equations \eqref{dbe} and \eqref{step}, the density does not evolve,
$n_j(t)=1$ when $j\leq 0$. Next, we define the mass in the positive
half space, $M_+(t)=\sum_{j=1}^\infty n_j(t)$, and note that summation
of \eqref{dbe} gives
\begin{equation}
\label{M+-eq}
\frac{dM_+}{dt}=n_0^2\, .
\end{equation}
Since $M_+(0)=0$ and $n_0(t)=1$, the mass equals time, $M_+=t$. This
fact, along with equation \eqref{wave} and the limiting behaviors
of the function $G(z)$, gives the velocity
\begin{equation}
\label{velocity}
v=1. 
\end{equation}
Hence, the propagation velocity, which is dictated by mass
conservation, agrees with the continuous value \cite{general}.

\begin{figure}[t]
\includegraphics[width=0.45\textwidth]{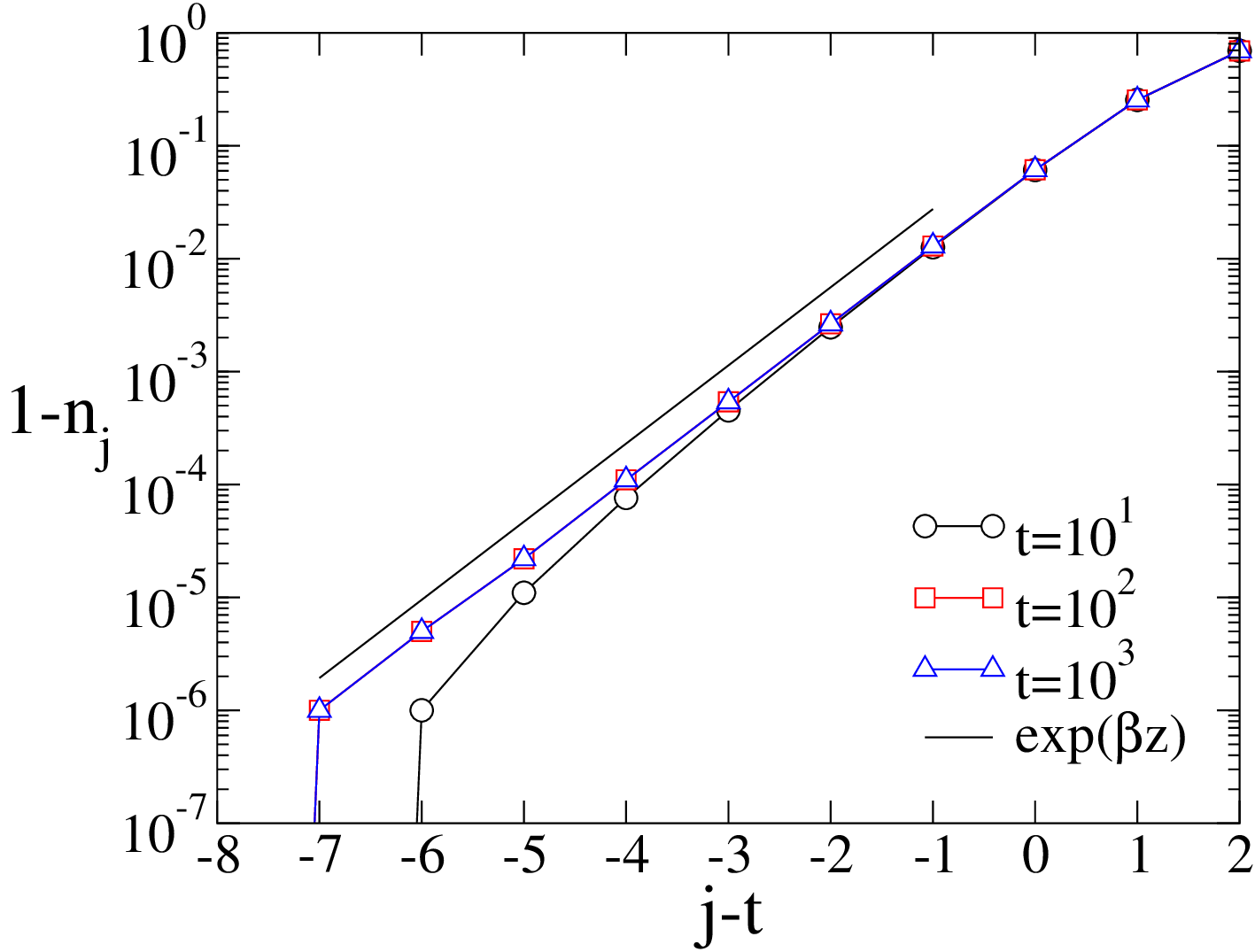}
\caption{The negative-$z$ tail of the function $G(z)$.  Shown is the
quantity $1-n_j$ versus $j-t$ at various times. The theoretical
prediction \eqref{tail1} with $\beta=1.59362$ is also shown for
reference.}
\label{fig-tail1}
\end{figure}

To characterize the shape of the traveling wave, we substitute the
form \eqref{wave} into the governing equation \eqref{dbe}. The
function $G(z)$ satisfies the nonlinear and nonlocal differential
equation
\begin{equation}
\label{G-eq}
G'(z)=G^2(z)-G^2(z-1),
\end{equation}
where prime denotes derivative with respect to $z$. We now use
asymptotic analysis to obtain the leading asymptotic behaviors of
$G(z)$ in the limits $z\to \pm\infty$.

Since $G(z)\to 1$ when $z\to -\infty$, we substitute
\hbox{$G(z)=1-\phi(z)$} into equation \eqref{G-eq}, and obtain a {\it
  linear} yet nonlocal equation for the correction function $\phi$,
\begin{equation}
\label{phi-eq}
\phi'(z)=2[\phi(z)-\phi(z-1)].
\end{equation}
Therefore, the correction decays exponentially,
\hbox{$\phi(z)\sim e^{\beta z}$}. By substituting this behavior into
\eqref{phi-eq}, we find the decay constant $\beta=1.59362$ as the
nontrivial root of the transcendental equation
$\beta=2(1-e^{-\beta})$. As shown in figure \ref{fig-tail1}, the
numerical results confirm that
\begin{equation}
\label{tail1}
1-G(z) \sim e^{\beta z}
\end{equation}
when $z\to -\infty$. This exponential behavior agrees, at least
qualitatively, with the corresponding behavior in the continuous case
\cite{special1}.

\begin{figure}[t]
\includegraphics[width=0.45\textwidth]{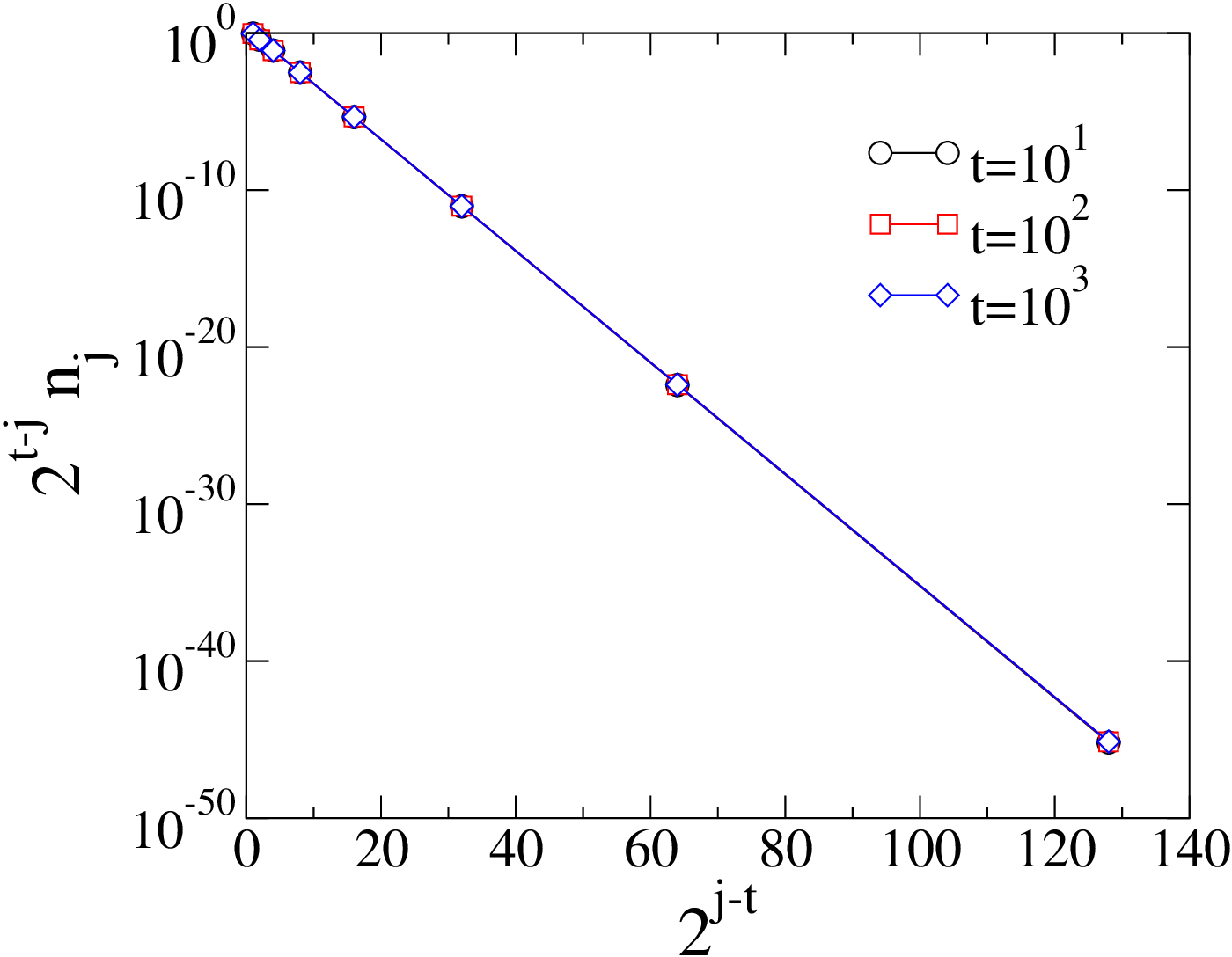}
\caption{The positive-$z$ tail of the traveling wave function $G(z)$.
Shown is $2^{j-t} n_j$ versus $2^{j-t}$ at three different times.}
\label{fig-tail2}
\end{figure}

In the complimentary $z\to \infty$ limit, we expect that $G(z)\to
0$. Now, the positive term in \eqref{G-eq} is negligible, and the
behavior is governed by the {\it nonlinear} and nonlocal equation
$G'(z)= -G^2(z-1)$. We use the WKB transformation \cite{bo}
$G(z)\simeq\Psi\exp(-\psi)$ and arrive at the linear relation
\begin{equation}
\psi(z)=2\psi(z-1).
\end{equation}
We thus obtain the exponential solution $\psi(z)=\gamma\,2^z$ and
further, the prefactor $\Psi(z)= (4\ln 2)\psi(z)$. The constant
$\gamma$ cannot be determined in the realm of asymptotic analysis;
numerically, we obtain $\gamma=0.818$.  Remarkably, the leading tail
of the traveling wave is extremely sharp, as it follows the unusual
double exponential decay (figure \ref{fig-tail2})
\begin{equation}
\label{tail2}
G(z) \sim 2^z e^{-\gamma\, 2^z}
\end{equation}
when $z\to\infty$. Hence, the front of the propagating wave does not
extend beyond a few lattice sites (see figure 
\ref{fig-traveling-wave}). In the continuous case the traveling wave
is specified by \hbox{$G_{\rm cont}(z)=\tfrac{1}{2}[1-\tanh(z/2\nu)]$}
and hence, the shape is symmetric as \hbox{$G_{\rm cont}(z)+G_{\rm
cont}(-z)=1$}. Therefore, the behavior in the discrete case differs in
two ways.  First, the shape of the traveling wave is
asymmetric. Second, the leading tail of the wave follows a
double-exponential decay, in contrast with the simple exponential
decay in the continuous case.

To gain further insight into the extremely sharp front \eqref{tail2},
we also study the continuous equation \eqref{cdbe}. We assume that the
solution approaches a traveling wave, \hbox{$n(x,t)\to g(x-vt)$},
with velocity $v=1$ (again, the velocity is set by the continuity
condition).  From Eq.~\eqref{cdbe}, the function $g(z)$ obeys the
ordinary differential equation
\begin{equation}
\label{g-eq-1}
g'-2gg' + (g g')'=0. 
\end{equation}
Integration of this equation is immediate, and using the limiting
behavior $g\to 0$ as $z\to\infty$, we have $g'=g-1$. We now invoke the
limiting behavior $g\to 1$ as $z\to -\infty$ and find
\begin{equation}
\label{g-sol}
g (z) = 
\begin{cases}
1 - e^{z-z_0}  & z\leq z_0,\\
0              & z\geq z_0.
\end{cases}
\end{equation}
This family of solutions is parameterized by $z_0$, a quantity that
depends on the details of the initial conditions. Hence, the traveling
wave is exponential {\em everywhere} behind the front which is located
at $t + z_0$. In particular, the negative-$z$ tail is analogous to
\eqref{tail1}.  Remarkably, the leading front of the traveling wave is
perfectly sharp \cite{ZK,B52,rh} as the function $g$ vanishes beyond
the front location!

We view the double exponential decay \eqref{tail2} as a discrete
analog of a perfectly sharp front. Moreover, the effective viscosity
$\nu\equiv n$ and the vanishing density ahead of the propagating front
are together responsible for the extremely sharp front \cite{general}.

\section*{Triangular Waves} 
Under the Burgers equation \eqref{be}, an initial condition with
compact support necessarily evolves into a triangular wave. Moreover,
the shape of the triangular wave is universal. Without loss of
generality, we consider the localized initial condition
\begin{equation}
\label{delta}
n_j(0)=\delta_{j,1}\,.
\end{equation}
As discussed above, according to the discrete equation \eqref{dbe},
the total mass is conserved. Moreover, there is no evolution in the
negative half-space, $n_j(t)=0$ for \hbox{$j\leq 0$}. Therefore, the
mass in the positive half space equals unity,
$M_+(t)=\sum_{j\geq 1} n_j(t)=1$.

\begin{figure}[t]
\includegraphics[width=0.45\textwidth]{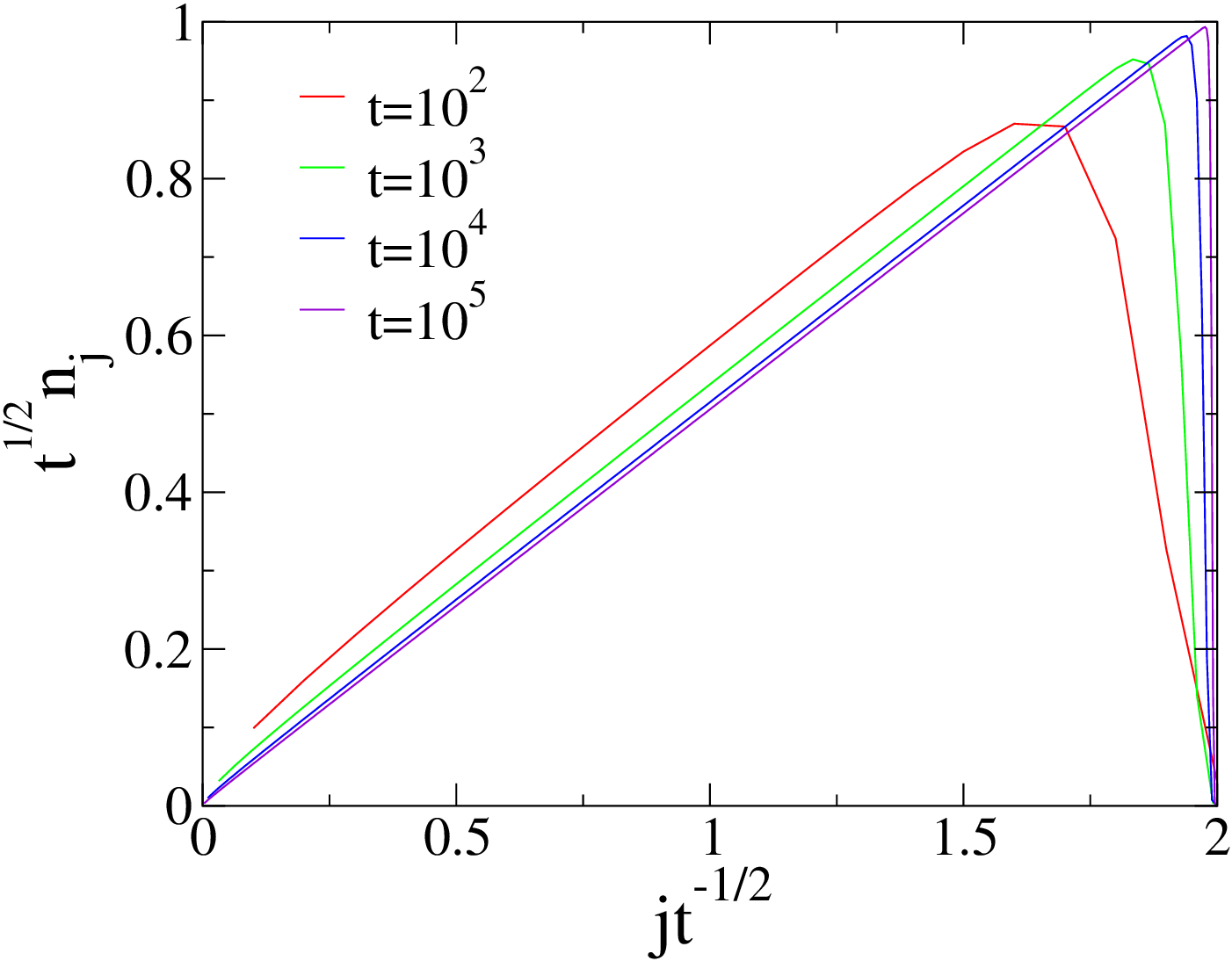}
\caption{The triangular wave. Shown is the function $t^{1/2}n_j(t)$
versus the scaling variable $jt^{-1/2}$ at different times.}
\label{fig-triangular-wave}
\end{figure}

Numerically, we integrate the discrete equation \eqref{dbe} starting
with the initial condition \eqref{delta} and find that the density
adheres to the scaling form (figure \ref{fig-triangular-wave})
\begin{equation}
\label{scaling-form}
n_j(t)\simeq \frac{1}{\sqrt{t}\,}F\left(\frac{j}{\sqrt{t}}\right).
\end{equation}
Further, the scaling function is triangular (figure
\ref{fig-triangular-wave})
\begin{equation}
\label{triangular}
F(x)=
\begin{cases}
0  & \quad x<0; \\
\frac{x}{2}& \quad0<x<2; \\
0 & \quad x>2.
\end{cases}
\end{equation}
Therefore, the discrete equation \eqref{dbe} captures the basic
features of the continuous equation \eqref{be}.

The scaling behavior \eqref{scaling-form}--\eqref{triangular} follows
from the inviscid Burgers equation \hbox{$\partial n/\partial
t+2n\,\partial n/\partial j=0$}. Indeed, mass conservation, in
conjunction with the inviscid Burgers equation, dictates the scaling
form \eqref{scaling-form} and the normalization $\int_0^\infty dx\,
F(x)=1$.  By substituting \eqref{scaling-form} into \eqref{be} and
ignoring the diffusion term, we see that function $F$ obeys
\begin{equation}
\frac{d}{dx}\left[F\left(F-\frac{x}{2}\right)\right]=0. 
\end{equation}
There are two solutions, $F=x/2$ and $F=0$, and consequently, the
scaling function is piecewise-linear as in
\eqref{triangular}. Finally, the extent of the nonzero region,
$0<x<2$, is set by the normalization condition.

We now analyze the solution of the discrete equation \eqref{dbe} in
the asymptotic regime. The quantity $n_1$ satisfies the closed
equation $dn_1/dt=-n_1^2$, and using the initial condition $n_1(0)=1$,
we have
\begin{equation}
\label{n1-sol}
n_1 =  \frac{1}{1+t}. 
\end{equation}
This density is inversely proportional to time, \hbox{$n_1\simeq
  t^{-1}$} in the limit $t\to\infty$. The structure of the rate
equations \eqref{dbe} suggests that in general, the densities $n_j$ are
inversely proportional to time,
\begin{equation}
\label{Aj-def}
n_j\simeq A_j\,t^{-1},
\end{equation}
when $t\to \infty$. Substitution of \eqref{Aj-def} into \eqref{dbe}
shows that this asymptotic behavior is compatible with the rate
equation, and furthermore, yields a quadratic recursion relation for
the prefactors,
\begin{equation}
\label{Aj-rec}
A_j^2 - A_j = A_{j-1}^2.
\end{equation}
Starting with $A_1=1$, we can iteratively obtain the prefactors from 
the explicit formula
\begin{equation}
\label{Aj-rec1}
A_j=\frac{1+\sqrt{1+4A_{j-1}^2}}{2}
\end{equation}
leading to $A_2=\frac{1+\sqrt{5}}{2}$, 
$A_3=\frac{1+\sqrt{7+2\sqrt{5}}}{2}$, etc. \cite{special2}.

We use a continuum approach to calculate the large-$j$ asymptotic
behavior of $A_j$. To obtain the leading asymptotic behavior, we
convert equation \eqref{Aj-rec} into the differential equation
$dA/dj=1/2$. Hence, to leading order, the coefficients are linear,
$A_j\simeq j/2$, and from the asymptotic behavior \eqref{Aj-def} one
has $n_j\simeq j/(2t)$. We thus recover the continuous behavior
indicated by \eqref{scaling-form} and $F(x)=x/2$.

There is, however, a correction to the leading asymptotic behavior. We
substitute $A_j=j/2+u_j$ into \eqref{Aj-rec} and find that the
correction $u_j$ obeys the recursion
\begin{equation}
(j-1)(u_j-u_{j-1})=\frac{1}{4}+u_{j-1}^2-u_j^2,
\end{equation}
and $u_1=1/2$.  When $j$ is large, the difference between the two
quadratic terms is negligible, and the continuous approach gives
\hbox{$du/dj=1/(4j)$}.  As a result, the prefactors include a
logarithmic correction
\begin{equation}
\label{Aj-sol}
A_j \simeq \frac{j}{2} + \frac{\ln j}{4} + C\, .
\end{equation}
The constant $C=0.32324$ is computed by numerical iteration of
\eqref{Aj-rec1}.

\begin{figure}[t]
\includegraphics[width=0.45\textwidth]{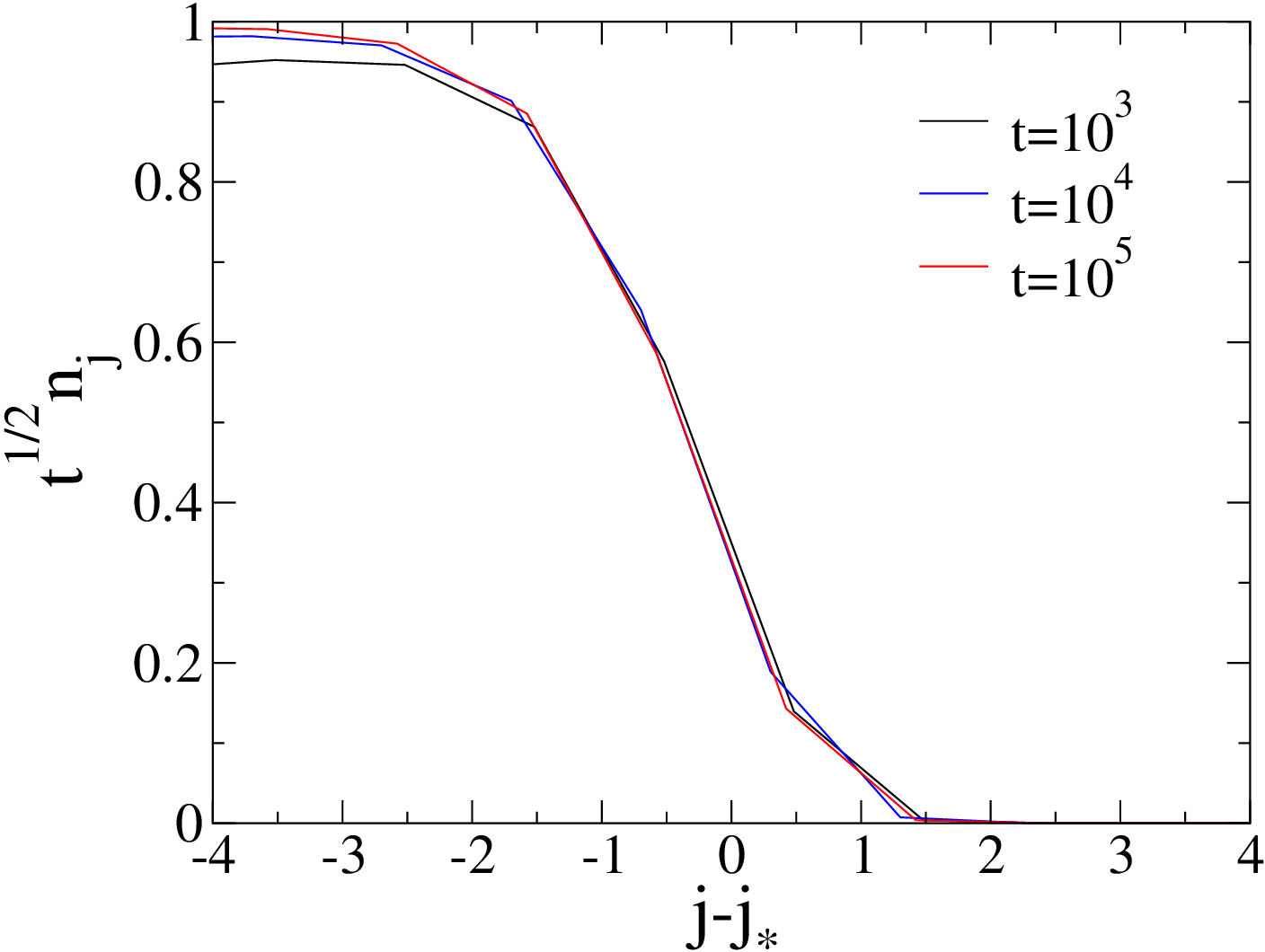}
\caption{The front of the triangular wave.  Shown is the function
  $t^{1/2}n_j(t)$ versus $j-j_*$ with $j_*$ given by \eqref{front} at
  different times.}
\label{fig-scaling-front}
\end{figure}

The logarithmic correction affects the location of the propagating
wave. If we denote the front location by $j_*$, then mass conservation
gives \hbox{$\sum_{j=1}^{j_*} n_j\simeq 1$}. Substitution of
\eqref{Aj-def} and \eqref{Aj-sol} into this sum yields
\begin{equation}
\label{front}
j_*\simeq 2\sqrt{t}-\tfrac{1}{4}\ln t\, ,
\end{equation}
by straightforward asymptotic analysis.  The leading term follows from
the continuous behaviors \eqref{scaling-form} and the extent of the
triangular wave specified by \eqref{triangular}. However, the
logarithmic correction, which follows from the coefficient
\eqref{Aj-sol} constitutes a departure from the continuous behavior.
We conclude that the triangular wave has the following form
\begin{equation}
\label{njt}
n_j(t)\simeq 
\begin{cases}
0 & j\leq 0,\\
A_jt^{-1}   &0<j\ll j_*,\\
0 & j\gg j_*.
\end{cases}
\end{equation}

We already established the front location $j_*$ and the maximal height
of the wave $n_*\sim t^{-1/2}$ which follows from \eqref{scaling-form}
and \eqref{triangular}. Thus, we anticipate the scaling behavior
\begin{equation}
\label{scaling-front}
n_j(t)\simeq \frac{1}{\sqrt{t}}\, G(j-j_*),
\end{equation}
in the finite neighborhood $j-j_*\sim {\cal O}(1)$.  Numerical
integration confirms this scaling behavior (figure
\ref{fig-scaling-front}).  The function $G(z)$ specifies the shape of
the front. By substituting \eqref{scaling-front} and \eqref{front}
into the rate equation \eqref{dbe}, we see that the scaling function
$G(z)$ satisfies \eqref{G-eq}. Hence, triangular waves and the
traveling waves have identical fronts!  This behavior is remarkable in
view of the two very different scaling forms \eqref{scaling-form} and
\eqref{scaling-front}.  In particular, the leading front is extremely
sharp and follows the double exponential decay \eqref{tail2}.

We also studied rarefaction waves by considering the complementary
step function
\begin{equation}
\label{step1}
n_j(0)=
\begin{cases}
0 & \quad j\leq 0,\\
1 & \quad j>0.
\end{cases}
\end{equation}
Again, there is no evolution in the negative half space. According to
the continuous equation \eqref{be}, the density obeys the scaling
behavior $n_j(t)\to F(j/t)$ with the very same $F(x)$ given by equation
\eqref{triangular}. Of course, there is a significant difference with
the triangular wave above as the extent of the wave is now linear in
time. In addition, there are diffusive boundary layers, for example
when $j-2t\sim {\cal O}(t^{1/2})$. Numerical integration of the rate
equation \eqref{dbe} shows that the scaling behavior and the diffusive
boundary layer both agree with the continuous case.  

In contrast with the traveling and triangular waves above, we can
replace the difference equation \eqref{dbe} with the differential
equation \eqref{be} even inside the front region because the size of
the boundary layer grows diffusively with time. However, the discrete
and the continuous equations are not equivalent when $j\sim {\cal
  O}(1)$ because of the vanishing viscosity.

Integrability is a remarkable feature of the Burgers equation: the
Cole-Hopf transformation $n=-\nu u_x/u$ turns the nonlinear equation
\eqref{be} into the ordinary diffusion equation $u_t=\nu u_{xx}$.  Our
discrete equation \eqref{dbe} is not integrable.  Interestingly,
however, the discrete equation
\begin{equation}
\label{dbe-modified}
\frac{dn_j}{dt}=n_j(n_j-n_{j-1}) 
\end{equation}
which has a purely quadratic right-hand-side and mimics the nonlinear
advection term in the Burgers equation, can be linearized. The
transformation \cite{br}
\begin{equation}
\label{trans}
n_j=\frac{q_j}{q_{j+1}}
\end{equation}
reduces the nonlinear equation \eqref{dbe-modified} into the linear
equation $dq_j/dt=q_{j-1}$.  Using this transformation, equation
\eqref{dbe-modified} can be solved exactly for the initial condition
\eqref{step1}. From the solution
$q_j=1+t+\tfrac{1}{2}t^2+\cdots+\tfrac{1}{(j-1)!}t^{j-1}$, the
rarefaction wave discussed above follows immediately.  However, the
discrete equation \eqref{dbe-modified} does not satisfy
continuity. Constructing a discrete analog of the Burgers equation
which satisfies the continuity condition, and in addition, can be
transformed into a linear equation is an interesting challenge.

In summary, we introduced a discrete nonlinear equation that captures
the key properties of the Burgers equation.  Intriguingly, the most
natural discrete analog of the Burgers equation is essentially unique:
matching the nonlinear advection term dictates that a viscosity that
is equal to the density, $\nu=n$.  The discrete equation satisfies the
continuity condition and consequently, the mass of an initially
localized ``bump'' is conserved. The mass conservation ensures that
the velocity of a traveling wave and the shape of a triangular wave
match the continuous behavior, to leading order. We also find that the
shape of a triangular wave is identical to the continuous case.

In contrast with the classic Burgers equation, the viscosity equals
the density. This discrepancy results in subtle departures from the
continuous behavior: The shape of the traveling wave is asymmetric,
and moreover, the front of the wave is extremely sharp with double
exponential tail. Further, the front of the triangular wave includes a
logarithmic-in-time correction to the leading behavior which is
diffusive.

\medskip
We gratefully acknowledge support for this research through DOE grant
DE-AC52-06NA25396.


\begin{thebibliography}{99}
\bibitem{general} In general, propagation from the state $n=a$ into
      the state $n=b<a$ occurs with the velocity $v=a+b$ and the
      function $G(z)$ is always asymmetric. When $b>0$, however, both
      tails of the function $G(z)$ are exponential.
\bibitem{special1} Qualitatively, the exponential behavior is
  consistent with asymptotic behavior of the density $n_1=\tanh t$.
\bibitem{special2} When $j=2$, the density $n_j$ can be obtained
      analytically \hbox{$n_2=A_2n_1 \!-\!
      (2A_2\!-\!1)\big[1+t+(1\!-\!A_2^{-1})(1+t)^{2A_2}\big]^{-1}$}.
\end{thebibliography}
\end{document}